\documentclass[12pt]{iopart}
\usepackage{iopams}
\usepackage{epsfig}

\begin{document}

\tolerance = 10000


\title{Absence of finite-temperature ballistic charge transport in the 1D half-filled Hubbard model}
\author{JMP Carmelo$^1$\,$^2$\,$^3$ and Shi-Jian Gu$^2$\,$^4$} 
\address{$^1$ Center of Physics, University of Minho, Campus Gualtar, 4710-057 Braga, Portugal\\
$^2$ Beijing Computational Science Research Center, Beijing 100084, China\\
$^3$ University of Gothenburg, 41296 Gothenburg, Sweden\\
$^4$ Department of Physics and Institute of Theoretical Physics, The Chinese University of Hong 
Kong, Hong Kong, China\\
E-mail: carmelo@fisica.uminho.pt}

\begin{abstract}
Finite-temperature $T>0$ transport properties of integrable and nonintegrable one-dimensional (1D)
many-particle quantum systems are rather different, showing in the metallic phases ballistic and diffusive behavior,
respectively. The repulsive 1D Hubbard model is an integrable system of wide physical interest. 
For electronic densities $n\neq1$ it is an ideal conductor, with ballistic charge transport for 
$T\geq 0$. In spite that it is solvable by the Bethe ansatz, at $n=1$ its $T>0$ transport 
properties are a collective-behavior issue that remains poorly understood. Here we combine that solution with symmetry to show 
that for on-site repulsion $U>0$ the charge stiffness $D (T)$ vanishes 
for $T>0$ in the thermodynamic limit. This absence of finite-temperature ballistic charge transport  
is an exact result that clarifies a long-standing open problem. 
\end{abstract}


\maketitle

\section{Introduction}
\label{Introduction}

The nature of the exotic transport properties of one-dimensional (1D) 
correlated electronic systems at finite temperature has been a problem of
long-standing interest \cite{CZP-95,ZP-96,ZNP-97,Kawa-98,
PDSC-00,PSZL-04,ANI-05,CGP-07,HPZ-11}.
The real part of the charge conductivity as a function of the frequency $\omega$ and
temperature $T$ has the form,
\begin{equation}
\sigma (\omega,T) = 2\pi\,D (T)\,\delta (\omega) +  \sigma_{reg} (\omega,T) \, .
\label{sigma}
\end{equation}
Here the charge stiffness $D(T)$ characterizes the response to a static field
and $\sigma_{reg} (\omega,T)$ describes the
absorption of light of frequency $\omega$. At $T>0$ the system can behave
as an ideal conductor with $D(T)>0$, a normal resistor with $D(T)=0$ and
$\sigma_0 = \lim_{\omega\rightarrow 0}\sigma_{reg} (\omega,T)>0$, and
an ideal insulator with $D(T)=\sigma_0 = 0$ \cite{CZP-95,ZP-96,PDSC-00,PSZL-04}.
1D normal conductors are typically correlated metallic nonintegrable electronic 
models, which show diffusive behavior such that the $T=0$ delta-fuction peak in 
the real part of the electrical conductivity broadens at $T>0$ into a Lorentzian Drude peak. 
An example of such nonintegrable systems is the 1D Hubbard-Peierls model \cite{polyace}. 
On the other hand, 1D ideal conductors are generally correlated metallic integrable electronic systems 
whose real part of the electrical conductivity shows a delta-fuction peak for $T\geq 0$. That
$D(T)>0$ for the latter systems, implies finite-temperature ballistic charge transport, 
the occurrence of an infinite set of conserving and commuting operators $\hat{Q}_j$
associated with the integrability preventing diffusive behavior for $T>0$ \cite{ZNP-97}.

The 1D Hubbard model is solvable using the Bethe ansatz (BA) \cite{Lieb,Takahashi,Martins97}.
This technique has been useful in the calculation of static properties \cite{Voit,1d-93-94,1d-RC-94}.
However, it has been difficult to apply to the study of transport at finite temperature.
The solvable 1D Hubbard model has $D(T)>0$ for electronic densities $n=N/N_a\neq 1$
and temperatures $T\geq 0$ \cite{PDSC-00,PSZL-04}. This result is consistent with an 
exact inequality involving the integrability conservation laws,
$D (T) > B (T) = 1/(2k_B T N_a)\sum_j \langle\hat J\hat{Q}_j\rangle^2/\langle\hat{Q}_j^2\rangle$.
Here $\langle ...\rangle$ stands for thermal averaging, $\hat J$ is the charge current operator,
and $B (T)>0$ for $n\neq 1$ provides a bound for the $D (T)$ value \cite{ZNP-97,Mazur,SPA-11}. 
However, one finds that $B (T)=0$ at $n=1$, so that the inequality is
inconclusive at half filling \cite{ZNP-97}. Indeed, the charge transport at $T>0$ is not well understood
for $N=N_a$ electrons and lattice sites.
For instance, whether in the thermodynamic limit and for on-site repulsion $U>0$
the charge stiffness $D(T)$ vanishes or is finite for $T>0$ and $n=1$ remains an open issue 
\cite{ZP-96,Kawa-98,PDSC-00,PSZL-04,CGP-07,HPZ-11}.

The authors of Ref. \cite{ZP-96} have conjectured that $D(T) =0$
at $n=1$ for $U/t>0$ and $N_a\rightarrow\infty$. Their analysis is based both on numerical results
for related integrable systems and on $D (0)=0$ exactly vanishing at $T=0$.
The studies of Ref. \cite{PDSC-00} rely on the BA solution and 
reach the exact result that $D(T) =0$ at $n=1$ to leading order in $t^2/U$
for $U/t\gg 1$ and $N_a\rightarrow\infty$. Here $t$ is the nearest-neighbor 
transfer integral. On the other hand, the investigations of
Ref. \cite{Kawa-98} also use the BA solution, yet  predict instead that
$D (T) >0$ at $n=1$ for $T>0$, $N_a\rightarrow\infty$, and $U/t>0$. However
and as discussed below in Sec. \ref{D-ome-reg}, their analysis has a fatal problem at $n=1$.

In this paper we fully clarify the above mentioned unsolved long-standing problem by 
showing that in the thermodynamic limit, $D(T)=0$ for $T>0$ at $n=1$ and $U/t>0$. 
Our result definitively establishes that for $U/t>0$ the half-filled 1D Hubbard model 
has no ballistic charge transport and thus is not an ideal conductor. 
Whether for $U/t>0$ and $T>0$ it is an ideal insulator or a normal resistor remains 
though an interesting open issue.

\section{The model global symmetry and energy eigenstates}
\label{model}

The 1D Hubbard model reads, 
\begin{equation}
\hat{H} = -t\sum_{\sigma}\sum_{j=1}^{N_a}\left[c_{j,\sigma}^{\dag}\,c_{j+1,\sigma} + 
{\rm h.c.}\right] + U\sum_{j=1}^{N_a}\hat{\rho}_{j,\uparrow}\hat{\rho}_{j,\downarrow} \, .
\label{H}
\end{equation}
Here $c_{j,\sigma}^{\dag}$ creates an electron of spin projection $\sigma$ at site $j$,
$\hat{\rho}_{j,\sigma}= (\hat{n}_{j,\sigma}-1/2)$, and $\hat{n}_{j,\sigma}=c_{j,\sigma}^{\dag}\,c_{j,\sigma}$. 
The states $\eta$-spin (and spin) and $\eta$-spin projection (and spin projection) are 
denoted by $S_{\eta}$ and $S_{\eta}^{z}$ (and $S_s$ and $S_s^{z}$), respectively. The 
$S_{\alpha}$ and $S_{\alpha}^{z}$ values of the lowest-weight states (LWSs) and 
highest-weight states (HWSs) of the $\eta$-spin and spin 
algebras are such that $S_{\alpha}= -S_{\alpha}^{z}$ and $S_{\alpha}=S_{\alpha}^{z}$, respectively. 
Here $\alpha =\eta$ for $\eta$-spin and $\alpha =s$ for spin. 

At $U=0$ the Hamiltonian of Eq. (\ref{H}) becomes that of a tight-binding model, whose energy eigenstates are as well
eigenstates of the current operator. One then trivially finds for $n=1$ and $N_a\rightarrow\infty$
that $\sigma_{reg} (\omega,T)=0$, $D (T)>0$, and max\,$D(T)=D(0)$, with $[D (0)-D(T)]\propto T^2>0$ 
and $D (T)\propto 1/T$ for low and high $T$, respectively. On the other hand, we find below 
that $D(T)=0$ for $U/t>0$. The $T\geq 0$ transition occurring at $U=U_c=0$ is controlled by the interplay 
of correlation effects with the emergence for $U/t>0$ 
of a hidden $U(1)$ symmetry beyond $SO(4)$ \cite{bipartite}. 
Indeed recently it was found that for $U/t\neq 0$ the global symmetry of the
Hubbard model on a bipartite lattice and thus in 1D is $[SO(4)\otimes U(1)]/Z_2$ \cite{bipartite}. 
The eigenvalue of the generator $2\tilde{S}_c^h$ of the hidden $U(1)$ symmetry beyond
$SO(4)$ is the number $2S_c^h$ of rotated-electron doubly plus unoccupied 
sites \cite{bipartite}. It is given by $2S_c^h=2[S_{\eta}+M']$ where $M'$ is the number of $\eta$-spin-neutral 
pairs of rotated-electron doubly and unoccupied sites 
\cite{bipartite}, which in 1D equals the BA number $M'$ of Ref. \cite{Takahashi}. 
The generator $2\tilde{S}_c^h$ does not commute with the charge current operator.

Importantly, the commutator $[\hat{H},2\tilde{S}_c^h]$ where $\hat{H}$ is the model Hamiltonian, Eq. (\ref{H}), 
is finite and vanishes at $U/t=0$ and for $U/t>0$, respectively.
Consistent, at $U/t=0$ the model global symmetry lacks the $U/t>0$ hidden $U(1)$ symmetry and
is instead $SO(4)\otimes Z_2$ \cite{bipartite}. Here the factor $Z_2$ refers to a discretely generated symmetry 
that is an exact symmetry of the $U/t=0$ Hamiltonian but changes the sign of the 
interaction Hamiltonian term in $U$ when $U>0$.
Taking the $U/t\rightarrow 0$ limit
of the $U/t>0$ energy eigenstates leads to eigenstates of $2\tilde{S}_c^h$ that
are different from the $U/t=0$ energy eigenstates, so that the problem is
nonperturbative. The $U/t$ dependence of the commutator of the model Hamiltonian with 
the hidden $U(1)$ symmetry generator then controls the corresponding phase
transition occurring at $U=U_c=0$ for $T\geq 0$. At $T=0$ it is the well known
Mott-Hubbard metal-insulator quantum phase transition \cite{Lieb}.

In the following we show that for $U/t>0$ the $D(T)$ value is determined by collective behavior
stemming from the interplay of the correlation effects with the algebra associated with the commutators
of the $\eta$-spin $SU(2)$ symmetry generators with several charge current operators.
For $U/t>0$ the model's BA solution has two alternative representations that refer to 
subspaces spanned either by the LWSs or HWSs of both $SU(2)$ symmetry algebras, respectively. 
In this paper we consider the LWS BA representation for which the numbers,
\begin{eqnarray}
n_{\eta} & = & S_{\eta} + S_{\eta}^z = 0,1,..., 2S_{\eta} \, ,
\nonumber \\
n_{s} & =  & S_{s} + S_{s}^z = 0,1,..., 2S_{s} \, ,
\label{n-eta.n-s}
\end{eqnarray}
vanish, where $S_{\eta}^z=-(N_a-N)/2$ and $S_{s}^z=-(N_{\uparrow}-N_{\downarrow})/2$. 
We call {\it Bethe states} the energy eigenstates contained in the BA subspace, which
are LWSs of both the $\eta$-spin and spin algebras. The spin non-LWSs are generated 
from such Bethe states by a transformation similar to that reported in the following for 
the $\eta$-spin non-LWSs, which involves the spin off-diagonal generators.
However, concerning the spin degrees of freedom our analysis considers general states, 
which may be spin LWSs or spin non-LWSs. Indeed the spin algebra plays no active role in our study.

For $U/t>0$ the $4^{N_a}$ energy eigenstates $\vert l_{\rm r},S_{\eta},S_{\eta}^z\rangle$ are as well
eigenstates of the hidden $U(1)$ symmetry generator $2\tilde{S}_c^h$, which counts the number of rotated-electron doubly 
plus unoccupied sites \cite{bipartite}. Within our notation, $l_{\rm r}$ stands 
for all quantum numbers beyond $S_{\eta}$ and $S_{\eta}^z$ needed to uniquely define a $U/t>0$ 
energy eigenstate, $\vert l_{\rm r},S_{\eta},S_{\eta}^z\rangle$. 
The $\eta$-spin non-LWSs are generated from the corresponding
$n_{\eta}=0$ $\eta$-spin LWS $\vert l_{\rm r},S_{\eta},-S_{\eta}\rangle$ as follows,
\begin{equation}
\vert l_{\rm r},S_{\eta},S_{\eta}^z\rangle = \vert l_{\rm r},S_{\eta},-S_{\eta}+n_{\eta}\rangle =
\frac{1}{\sqrt{{\cal{C}}_{\eta}}}({\hat{S}}^{+}_{\eta})^{n_{\eta}}\vert l_{\rm r},S_{\eta},-S_{\eta}\rangle \, .
\label{Gstate-BAstate}
\end{equation}
Here $n_{\eta} = 1,...,2S_{\eta}$,
\begin{equation}
{\cal{C}}_{\eta} = \langle l_{\rm r},S_{\eta},-S_{\eta}\vert ({\hat{S}}_{\eta}^{-})^{n_{\eta}}({\hat{S}}^{+}_{\eta})^{n_{\eta}}\vert l_{\rm r},S_{\eta},-S_{\eta}\rangle
= [n_{\eta}!]\prod_{j=1}^{n_{\eta}}[\,2S_{\eta}+1-j\,] \, , 
\label{Calpha}
\end{equation}
is a normalization constant and the $\eta$-spin generators read,
\begin{eqnarray}
{\hat{S}}^{+}_{\eta} & = & \sum_{j=1}^{N_a}(-1)^j\,c_{j,\downarrow}^{\dag}\,
c_{j,\uparrow}^{\dag} \, ; \hspace{0.25cm}
{\hat{S}}_{\eta}^{-} =({\hat{S}}_{\eta}^{+})^{\dag}
\, ; \hspace{0.25cm}
{\hat{S}}_{\eta}^{z} = {1\over 2}\sum_{j=1}^{N_a}(\hat{\rho}_{j,\uparrow}+\hat{\rho}_{j,\downarrow}) \, ,
\nonumber \\
(\hat{\vec{S}}_{\eta})^2 & = & (\hat{S}_{\eta}^{z})^2 + {1\over 2}(\hat{S}_{\eta}^{+}\hat{S}_{\eta}^{-}
+ \hat{S}_{\eta}^{-}\hat{S}_{\eta}^{+}) \, ,
\label{S+S+}
\end{eqnarray}
where we have also provided the diagonal generator and $(\hat{\vec{S}}_{\eta})^2$ expressions.
Importantly, the half-filling energy eigenstates, which are those of most interest for our study, refer 
in Eq. (\ref{Gstate-BAstate}) to $n_{\eta}=S_{\eta}$. 

Except for a constant pre-factor, the charge current operator $\hat{J}^{\rho}$ equals 
the $z$-axis $\eta$-spin current operator $\hat{J}^{\sigma_{z}^{\eta}}$,
\begin{eqnarray}
\hat{J}^{\rho} & = & (e)\,\hat{J} \, ; \hspace{0.50cm} \hat{J}^{\sigma_{z}^{\eta}} = (1/2)\,\hat{J} \, ,
\nonumber \\
\hat{J} & = & -i\,t\sum_{\sigma}\sum_{j=1}^{N_a}\left[c_{j,\sigma}^{\dag}\,c_{j+1,\sigma} - 
c_{j+1,\sigma}^{\dag}\,c_{j,\sigma}\right] \, ,
\label{c-s-currents}
\end{eqnarray}
where $e$ denotes the electronic charge. 

Our main goal is to calculate the charge stiffness $D(T)$ for $U/t>0$ and $S_{\eta}^z =0$, 
which in the thermodynamic limit involves only current expectation values and can be written 
in terms of a sum over the $S_{\eta}^z =0$ half-filling states, Eq. (\ref{Gstate-BAstate}), as \cite{ZNP-97,PDSC-00},
\begin{equation}
2\pi D (T) = 
{\pi\over k_B T N_a}\sum_{l_{\rm r}} \sum_{S_{\eta}=0,1,2,...,N_a/2} p_{l,S_{\eta}} \vert\langle l_{\rm r},S_{\eta},0\vert\hat{J}
\vert l_{\rm r},S_{\eta},0\rangle\vert^2 \, .
\label{D-T}
\end{equation}
Here $p_{l_{\rm r},S_{\eta}}=e^{-\epsilon_{l_{\rm r},S_{\eta}}/k_BT}/Z$ is the usual Boltzmann weight 
and the partition function reads $Z = \sum_{l_{\rm r},S_{\eta}}  e^{-\epsilon_{l_{\rm r},S_{\eta}}/k_BT}$.
 
\section{current matrix elements and expectation values}
\label{matr-el-expec-val}

The following commutators play a major role in our study,
\begin{eqnarray}
\left[\hat{J},\hat{S}_{\eta}^{z}\right] & = & 0
\, ; \hspace{0.50cm}
\left[\hat{J},(\hat{\vec{S}}_{\eta})^2\right]  = \hat{J}^{+}\hat{S}_{\eta}^{-} - \hat{S}_{\eta}^{+}\hat{J}^{-}
\nonumber \\
\left[\hat{J},\hat{S}_{\eta}^{\pm}\right] & = &
\left[\hat{S}_{\eta}^{z},\hat{J}^{\pm}\right] = \pm \hat{J}^{\pm} 
\, ; \hspace{0.50cm}
\left[\hat{J}^{\pm},\hat{S}_{\eta}^{\mp}\right] = \pm 2\hat{J} \, ,
\label{comm-currents}
\end{eqnarray}
where $\hat{J}^{\pm}$ denotes the following current operators related to the $\eta$-spin
$SU(2)$ symmetry algebra,
\begin{equation}
\hat{J}^{+} = i\,2t\sum_{j=1}^{N_a}(-1)^j\left[c_{j,\downarrow}^{\dag}\,c_{j+1,\uparrow}^{\dag} +
c_{j+1,\downarrow}^{\dag}\,c_{j,\uparrow}^{\dag}\right]  
\, ; \hspace{0.50cm}
\hat{J}^{-}=(\hat{J}^{+})^{\dag} \, .
\label{+-currents}
\end{equation}

The $S_{\eta}>0$ metallic $\eta$-spin LWSs $\vert l_{\rm r} ,S_{\eta},-S_{\eta}\rangle$ and half-filling
simultaneously $\eta$-spin LWSs and HWSs  $\vert l_{\rm r} ,0,0\rangle$ used
in our operator algebra manipulations obey the following well-known transformation laws,
\begin{eqnarray}
\hat{S}_{\eta}^{-}\vert l_{\rm r} ,S_{\eta},-S_{\eta}\rangle & = & 0 \, ,
\nonumber \\
\hat{S}_{\eta}^{+}\vert l_{\rm r} ,0,0\rangle & = & \hat{S}_{\eta}^{-}\vert l_{\rm r} ,0,0\rangle = 0 \, ,
\label{SS0}
\end{eqnarray}
which trivially follow from the $\eta$-spin $SU(2)$ symmetry algebra. 

In order to evaluate the current expectation values $\langle l_{\rm r} ,S_{\eta},0\vert\hat{J}\vert l_{\rm r} ,S_{\eta},0\rangle$ that
appear in the charge-stiffness expression, Eq. (\ref{D-T}), in the following we consider a more
general problem: That of finding from operator-algebra symmetry alone some of the general current matrix elements,
\begin{eqnarray}
\langle l_{\rm r} ,S_{\eta},S_{\eta}^z\vert\hat{J}\vert l_{\rm r},S_{\eta}',S_{\eta}^{z}\rangle & = & {1\over \sqrt{{\cal{C}}_{\eta}{\cal{C}}_{\eta}'}}
\nonumber \\
& \times & \langle l_{\rm r} ,S_{\eta},-S_{\eta} \vert ({\hat{S}}_{\eta}^-)^{n_{\eta}}\hat{J} 
({\hat{S}}^+_{\eta})^{n_{\eta}'}\vert l_{\rm r},S_{\eta}',-S_{\eta}'\rangle \, ,
\label{J-non-LWS}
\end{eqnarray}
that vanish. Here $n_{\eta} = S_{\eta} + S_{\eta}^z$, $n_{\eta}' = S_{\eta}' + S_{\eta}^z$, the normalization constants are
given in Eq. (\ref{Calpha}), and we have accounted for the vanishing of the commutator $[\hat{J},\hat{S}_{\eta}^{z}] = 0$,
Eq. (\ref{comm-currents}), so that the current operator only connects states with the same $S_{\eta}^z$ 
value. For $l{\rm r}=l{\rm r}'$, $S_{\eta}=S_{\eta}'$, and $S_{\eta}^z=0$ Eq. (\ref{J-non-LWS}) 
refers to the current expectation values in Eq. (\ref{D-T}).

To double check our results on the current expectation values, we find their value from limiting cases of
two different classes of current matrix elements: (a) matrix elements
$\langle l_{\rm r} ,S_{\eta},0\vert\hat{J}\vert l_{\rm r},S_{\eta}',0\rangle$ between $S_{\eta}^{z}=0$
half-filling states with arbitrary $S_{\eta}$ and $S_{\eta}'$ values, respectively, and (b) matrix elements
$\langle l_{\rm r} ,S_{\eta},S_{\eta}^z\vert\hat{J}\vert l_{\rm r},S_{\eta},S_{\eta}^{z}\rangle$
between states with the same $S_{\eta}>0$ and $S_{\eta}^{z}$ arbitrary values. While the current matrix elements
of type (a) connect only half-filling states those of type (b) may connect $S_{\eta}^{z}>0$ metallic states.

By combining the systematic use of the commutators given in Eq. (\ref{comm-currents}) with the
transformation laws of Eq. (\ref{SS0}), we reach the following general useful result concerning
the current matrix elements of type (a),
\begin{equation}
\langle  l_{\rm r} ,S_{\eta},0\vert\hat{J}\vert  l_{\rm r},S_{\eta}+\delta S_{\eta},0\rangle = 0 \, ,
\hspace{0.5cm} \delta S_{\eta} \neq \pm 1 \, .
\label{J-non-LWS}
\end{equation}
For half-filling states whose generation from metallic $\eta$-spin LWSs involves small $n_{\eta}=(S_{\eta}-S_{\eta}^z)$ 
values, the calculations are straightforward. They become lengthly as the $n_{\eta}$ value increases, yet remain 
straightforward.

Furthermore, by use of similar techniques we find after a suitable operator algebra involving commutator 
manipulations and state transformations the following relation between current matrix elements of type (b),
\begin{eqnarray}
\langle l_{\rm r},S_{\eta},S_{\eta}^z\vert\hat{J}\vert l_{\rm r}',S_{\eta},S_{\eta}^z\rangle =
C (2S_{\eta},n_{\eta})\,\langle l_{\rm r},S_{\eta},-S_{\eta}\vert \hat{J}\vert l_{\rm r}',S_{\eta},-S_{\eta}\rangle \, , 
\label{currents-gen}
\end{eqnarray}
where $S_{\eta}^z = -S_{\eta} + n_{\eta}$ and $n_{\eta} = 1,...,2S_{\eta}$.
The coefficient $C (l,\tilde{l})$ appearing here is such that,
\begin{eqnarray}
C (l,\tilde{l}) & = & -C (l,l -\tilde{l}) \, , \hspace{0.35cm} \tilde{l} \leq l/2 \, ,
\nonumber \\
C (l,l/2) & = & 0 \hspace{0.15cm}{\rm for}\hspace{0.15cm}l/2\hspace{0.15cm}{\rm integer} \, .
\label{rela-exp-s-currents}
\end{eqnarray}
The result $C (l,l/2) = 0$ follows from the first equality for $\tilde{l}= l/2$, where
we have denoted $2S_{\eta}$ and $n_{\eta}$ by $l$ and $\tilde{l}$, respectively.

First, it follows from Eq. (\ref{J-non-LWS}) for $l_{\rm r}=l_{\rm r}'$ and $\delta S_{\eta}=0$
that the expectation values of all half-filling energy eigenstates $\vert  l_{\rm r},S_{\eta},0\rangle$ vanish,
$\langle  l_{\rm r} ,S_{\eta},0\vert\hat{J}\vert  l_{\rm r},S_{\eta},0\rangle = 0$.
Second, for $S_{\eta}>0$ such half-filling states have numbers $S_{\eta} = n_{\eta} = 1,2,...,N_a/2$, so that 
$C (2S_{\eta},n_{\eta})=C (2S_{\eta},S_{\eta}) =0$, as given in Eq. (\ref{rela-exp-s-currents}).
Consistent with the vanishing current expectation values of all $S_{\eta}\geq 0$ found from Eq. (\ref{J-non-LWS}),
it follows from Eq. (\ref{currents-gen}) that the current expectation
value of $S_{\eta}>0$ half-filling states $\vert  l_{\rm r},S_{\eta},0\rangle$ vanishes.

We have then confirmed that the analysis of the two classes of current matrix elements 
leads to the same result, that for $U/t>0$ the charge current expectation values
$\langle l_{\rm r},S_{\eta},0\vert\hat{J}\vert l_{\rm r},S_{\eta},0\rangle$
vanish for all $n=1$ energy eigenstates. In contrast, the current expectation values
$\langle l_{\rm r},S_{\eta},S_{\eta}^z\vert\hat{J}\vert l_{\rm r},S_{\eta},S_{\eta}^z\rangle$ of $S_{\eta}^z\neq 0$
metallic states are in general finite \cite{PCCS-97,GPC-07}.
\begin{table}
\begin{tabular}{|c|c|c|c|c|c|c|c|c|} \hline
$S_{\eta}\backslash n_{\eta}$ & $0$ & $1$ & $2$ & $3$ & $4$ & $5$ & $6$ & $7$  \\
\hline
1/2 & 1 & -1 & - & - & - & - & - & - \\
\hline
1 & 1 & 0 & -1 & - & - & - & - & - \\
\hline
3/2 & 1 & 1/3 & -1/3 & -1 & - & - & - & - \\
\hline
2 & 1 & 1/2 & 0 & -1/2 & -1 & - & - & - \\
\hline
5/2 & 1 & 3/5 & 1/5 & -1/5 & -3/5 & -1 & - & - \\
\hline
3 & 1 & 2/3 & 1/3 & 0 & -1/3 & -2/3 & -1 & - \\
\hline
7/2 & 1 & 5/7 & 9/16 & 1/7 & -1/7 & -9/6 & -5/7 & -1 \\
\hline
\end{tabular}
\caption{The coefficient $C (2S_{\eta},n_{\eta})$ on the right-hand side of Eq. (\ref{currents-gen})
for the $\eta$-spin-tower states of $\eta$-spin up to $S_{\eta}=7/2$. At half filling one has that $S_{\eta}=0,1,2,3$ 
is an integer and $S_{\eta}^z =0$, so that $n_{\eta}=S_{\eta}$ and $C (2S_{\eta},S_{\eta})=0$.}
\label{table-j}
\end{table}

Only the coefficient $C (2S_{\eta},n_{\eta})=C (2S_{\eta},S_{\eta}) =0$ in Eqs. (\ref{currents-gen}) and (\ref{rela-exp-s-currents})
is needed for our $n=1$ study. A general expression of that coefficient which applies to
$n_{\eta}=0,1,2,3$ and any $\eta$-spin value $S_{\eta}\leq n_{\eta}/2$ is,
\begin{eqnarray}
C (l,\tilde{l}) & =  & {1\over [\tilde{l}!]\prod_{j=1}^{\tilde{l}}[\,l+1-j\,]}
[\prod_{j=1}^{\tilde{l}}[\,j l-2^j\,] - (\tilde{l}-1)(2[(l+1)-(\tilde{l}-1)])^{\tilde{l}-1}
\nonumber \\
& - & (1-\delta_{\tilde{l},1})(l\tilde{l}-2^{\tilde{l}}) (\tilde{l}-2)(2[(l+1)
- (\tilde{l}-2)])^{\tilde{l}-2}] \, , \hspace{0.35cm}
\tilde{l} = 1,2,3 \, , \hspace{0.35cm} l \geq \tilde{l} \, , 
\label{exp-s-currents-n1-3}
\end{eqnarray}
where $l \equiv 2S_{\eta}$ and $\tilde{l} \equiv n_{\eta}$. For $n_{\eta}>3$ the $C (2S_{\eta},n_{\eta})$ expression becomes too 
cumbersome for $S_{\eta}\neq n_{\eta}$ metallic states and vanishes for half-filling $S_{\eta}= n_{\eta}$ states.
Combining the expression of Eq. (\ref{exp-s-currents-n1-3}) with the relation $C (l,\tilde{l}) = -C (l,l -\tilde{l})$ provided in Eq. 
(\ref{rela-exp-s-currents}) for $\tilde{l} \leq l/2$, we have calculated the coefficient $C (2S_{\eta},n_{\eta})$ of 
all states with $\eta$-spin $S_{\eta}\leq 7/2$, whose values are given in Table \ref{table-j}. 

\section{The charge stiffness and regular conductivity at half filling}
\label{D-ome-reg}

Our above result that the current expectation values of all $U/t>0$ half-filling energy eigenstates
$\vert l_{\rm r},S_{\eta},0\rangle$ vanish implies according to Eq. (\ref{D-T}) that the
charge stiffness $D(T)$  vanishes in the thermodynamic limit. Hence we have just showed that at half filling 
it vanishes for $U/t>0$ and $T>0$ in the thermodynamic limit, whereas $D(T)>0$ at $U/t=0$.
This is our main result, which clarifies a long-standing open problem. Note however that the conductivity 
sum rule $\int d\omega\,\sigma (\omega,T)$ remains invariant under
the transition occurring at $U=U_c=0$ for all temperatures. Indeed, we find that
$2\pi\,D (T)\vert_{U/t=0} = \lim_{U/t\rightarrow 0}\int d\omega\,\sigma_{reg} (\omega,T)>0$ and
$\lim_{U/t\rightarrow 0}2\pi\,D (T) = \left[\int d\omega\,\sigma_{reg} (\omega,T)\right]\vert_{U/t=0}=0$.

We emphasize that our exact result that $D(T)$ vanishes at $n=1$ for $U/t>0$ and 
$T\geq 0$ does not apply to the model on a finite 1D lattice. For it the charge stiffness 
expression has additional terms, beyond those given in Eq. (\ref{D-T}), which vanish in
the present thermodynamic limit \cite{ZNP-97}. Such extra terms involve current matrix elements 
between pairs of degenerate energy eigenstates.
Moreover, our exact results disagree with the prediction of Ref. \cite{Kawa-98} that
$D(T)$ should be finite in the thermodynamic limit for $U/t>0$ and $n=1$. 
That prediction error stems from some of the separate integrals of the individual summands occurring 
in the integrands of Eq. (25) of Ref. \cite{Kawa-98}, which diverge at $n=1$. That turns out to be a fatal problem, 
similar to that of some of the integrands of Eqs. (24) and (25) 
of Ref. \cite{ANI-05} for a related BA solvable model, as was discussed and recognized in that reference. 
On the other hand, the studies of Ref. \cite{CGP-07} did not calculate explicitly the charge currents
carried by $S_{\eta}^z=0$ states with $S_{\eta}>0$ and assumed those to be finite, alike for
the metallic states of the same $\eta$-spin-$S_{\eta}$ tower, yet they vanish.

Our results allow two possible scenarios for the 1D half-filled Hubbard model phase at a given finite temperature $T$: Either 
the model behaves as a normal resistor with $D(T)=0$ and $\sigma_0 = \lim_{\omega\rightarrow 0}\sigma_{reg} (\omega,T)>0$
or as an ideal insulator with $D(T)=\sigma_0 = 0$.
The use of Eq. (\ref{J-non-LWS}) allows the simplification of the standard linear-response 
theory expression of $\sigma_{reg} (\omega,T)$ to,
\begin{eqnarray}
& & \sigma_{reg} (\omega,T) = {\pi\left(1-e^{-\omega/(k_B\,T)}\right)\over N_a\,\omega}
\sum_{l_{\rm r}} \sum_{S_{\eta}=0,1,2,...,N_a/2} p_{l_{\rm r},S_{\eta}} \sum_{j=\pm 1}\Theta (S_{\eta}+j)
\nonumber \\
& \times &\sum_{l_{\rm r}',(\epsilon_{l_{\rm r}',S_{\eta}+j}\neq \epsilon_{l_{\rm r},S_{\eta}})}
\vert\langle l_{\rm r},S_{\eta},0\vert\hat{J}\vert l_{\rm r}',S_{\eta}+j,0\rangle\vert^2
\delta (\omega -\epsilon_{l',S_{\eta}+j}+\epsilon_{l,S_{\eta}}) \, .
\label{sigma-reg-SS}
\end{eqnarray}
Here $\Theta (x)=1$ for $x\geq 0$ and $\Theta (x)=0$ for $x< 0$.

For $U/t>0$ the exact ground state of the half-filled Hubbard model at zero chemical potential and
zero spin density, which here we denote by $\vert GS,0,0\rangle$,
is a $\eta$-spin singlet, $S_{\eta}=0$, with $M'=0$ \cite{Lieb,Takahashi}, so that it is an eigenstate of the 
hidden $U(1)$ symmetry generator with
eigenvalue $2S_c^h = 2[S_{\eta}+M'] = 0$. Moreover, the exact minimum energy for transitions from that ground state 
to two-electron charge and $\eta$-spin excited states with $2S_c^h=2[S_{\eta}+M']>0$ is
min\,$\Delta_{D_{rot}} = 2\Delta_{MH}\,[S_{\eta}+M']$. Here $2\Delta_{MH}$ is the
Mott-Hubbard gap, which at zero spin density behaves as $\sim {8\over \pi}\sqrt{t\,U}\,e^{-2\pi \left({t\over U}\right)}$ for $U/t\ll 1$
and $\sim (U - 4t)$ for $U/t\gg 1$ \cite{Lieb}. The minimum excitation energy, min\,$\epsilon_{l',1}-\epsilon_{GS,0}=2\Delta_{MH}$,  
relative to the $S_{\eta}=0$ ground state whose matrix element $\langle GS,0,0\vert\hat{J}\vert l_{\rm r}',1,0\rangle$
in Eq. (\ref{sigma-reg-SS}) does not vanish refers to excited energy eigenstates $\vert l_{\rm r}',1,0\rangle$ with $S_{\eta}=1$ 
and $2S_c^h=2$. Hence at $T=0$ we find that $\sigma_{reg} (\omega,0)= 0$ for $\omega < 2\Delta_{MH}$.
This confirms that the real part of the conductivity vanishes for the $T=0$ Mott-Hubbard insulator for
energies smaller than the Mott-Hubbard gap. 
\begin{figure}
\includegraphics[scale=0.30]{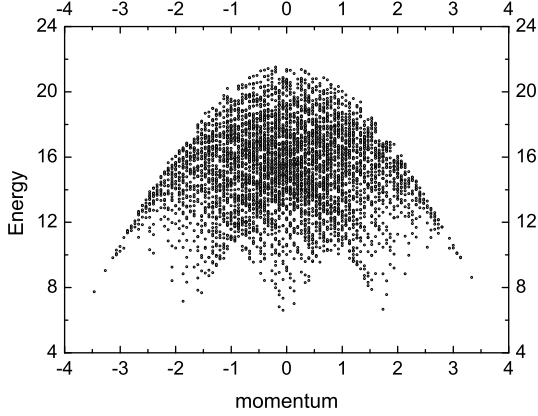}
\caption{\label{fig1}The degenerate energy spectrum of several types 
of $2S_c^h = 2[S_{\eta}+M'] = 4$ states for $ U/t=6$ and $N_a=30$. This includes the
$S_{\eta}=2;M'=0$ states, $S_{\eta}=1;M'=1$ states, and $S_{\eta}=0;M'=2$ 
states considered in the text.}
\end{figure}

To characterize possible $T>0$ transitions for which $\epsilon_{l_{\rm r}',S_{\eta}\pm1}-\epsilon_{l_{\rm r},S_{\eta}}\rightarrow 0$
in Eq. (\ref{sigma-reg-SS}), it is convenient to replace the quantum number $l_{\rm r}$ in 
$\vert l_{\rm r},S_{\eta},S_{\eta}^z\rangle$ by two quantum numbers, $m_{\rm r},M'$, so that 
$\vert m_{\rm r}, M',S_{\eta},S_{\eta}^z\rangle\equiv \vert l_{\rm r},S_{\eta},S_{\eta}^z\rangle$. 
Here $m_{\rm r}$ stands now for all quantum numbers beyond $M'$, $S_{\eta}$, and $S_{\eta}^z$ needed 
to uniquely define the $U/t>0$ energy eigenstate. From analysis of the half-filling energy spectra
obtained by combining the BA solution with symmetry,
we then find that $\epsilon_{m_{\rm r}',M'-1,S_{\eta}}-\epsilon_{m_{\rm r},M',S_{\eta}-1}\rightarrow 0$ for pairs of states with the
same hidden $U(1)$ symmetry generator eigenvalue $2S_c^h = 2[S_{\eta}+M']$ and
suitable $m_{\rm r}$ and $m_{\rm r}'$ values. Specifically, provided that the matrix elements of the following form
are finite, 
\begin{eqnarray}
& & \langle m_{\rm r},M',S_{\eta}-1,0\vert\hat{J}
\vert m_{\rm r}', M'-1,S_{\eta},0\rangle\vert_{S_{\eta}=M'} = {1\over\sqrt{{\cal{C}}_{\eta}{\cal{C}}_{\eta}'}}
\nonumber \\
& \times & \langle m_{\rm r},M',S_{\eta}-1,-S_{\eta}+1\vert
({\hat{S}}_{s}^-)^{M'-1}\hat{J}
({\hat{S}}^+_{\eta})^{M'}
\vert m_{\rm r}', M'-1,S_{\eta},-S_{\eta}\rangle\vert_{S_{\eta}=M'}
\nonumber \\
& = & {1\over\sqrt{C_{M'}}}\langle m_{\rm r},M',S_{\eta}-1,-S_{\eta}+1\vert\hat{J}^+
\vert m_{\rm r}', M'-1,S_{\eta},-S_{\eta}\rangle\vert_{S_{\eta}=M'} \, ,
\label{cond-mel}
\end{eqnarray}
where $M'=1,2,...$ and the two states are such that $\epsilon_{m_{\rm r}',M'-1,S_{\eta}}-\epsilon_{m_{\rm r},M',S_{\eta}-1}\rightarrow 0$, 
then $\sigma_0 = \lim_{\omega\rightarrow 0}\sigma_{reg} (\omega,T)$
would be finite for $T>0$. Note that the states $\vert m_{\rm r}', M'-1,S_{\eta},-S_{\eta}\rangle$
and $\vert m_{\rm r},M',S_{\eta}-1,-S_{\eta}+1\rangle$ connected in Eq. (\ref{cond-mel})
by the two-electron current operator $\hat{J}^+$, Eq. (\ref{+-currents}),
have $N=N_a-2M'$ and $N+2=N_a-2M' +2$ electrons, respectively. 

In case that such matrix elements were
finite, their absolute value would decrease rapidly upon increasing $M'$ and a large fraction of
the weight would be generated by the $M'=1$ transition. (The constant $C_{M'}$ in the
last expression given in Eq. (\ref{cond-mel})
reads $C_1 = 2$ and $C_2 = 3$ for $M'=1$ and $M'=2$, respectively.) 
Unfortunately, we could not evaluate such matrix elements.
The numerical results of Ref. \cite{PSZL-04} refer to finite systems and provide 
some evidence that $\sigma_0$ could be finite in the thermodynamic limit for $U/t>0$ 
and $T\rightarrow\infty$. Nonetheless an ultimate prove that for $U/t>0$ the
conductivity $\sigma_0$ vanishes or is finite remains lacking. 
\begin{figure}
\includegraphics[scale=0.30]{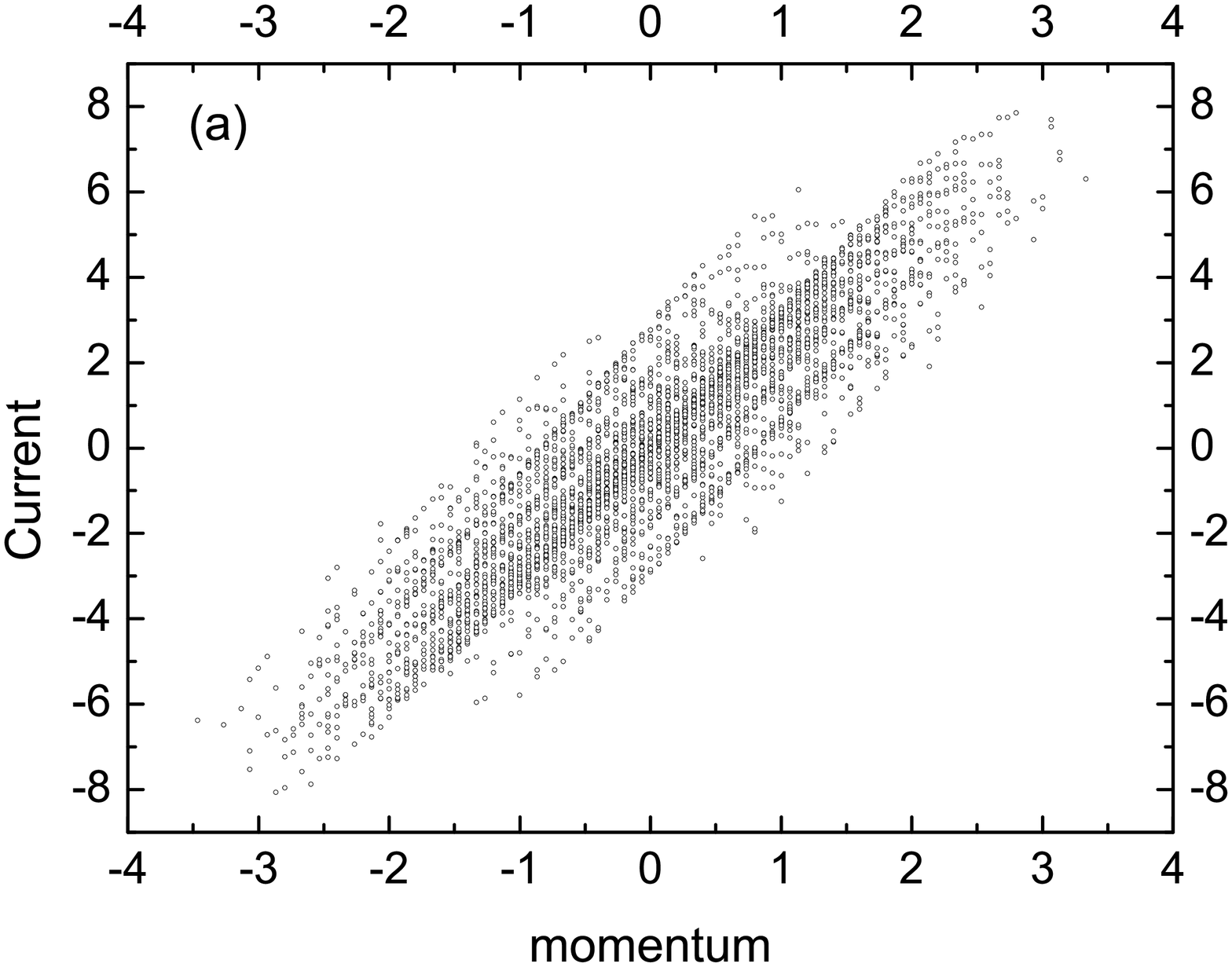}
\includegraphics[scale=0.30]{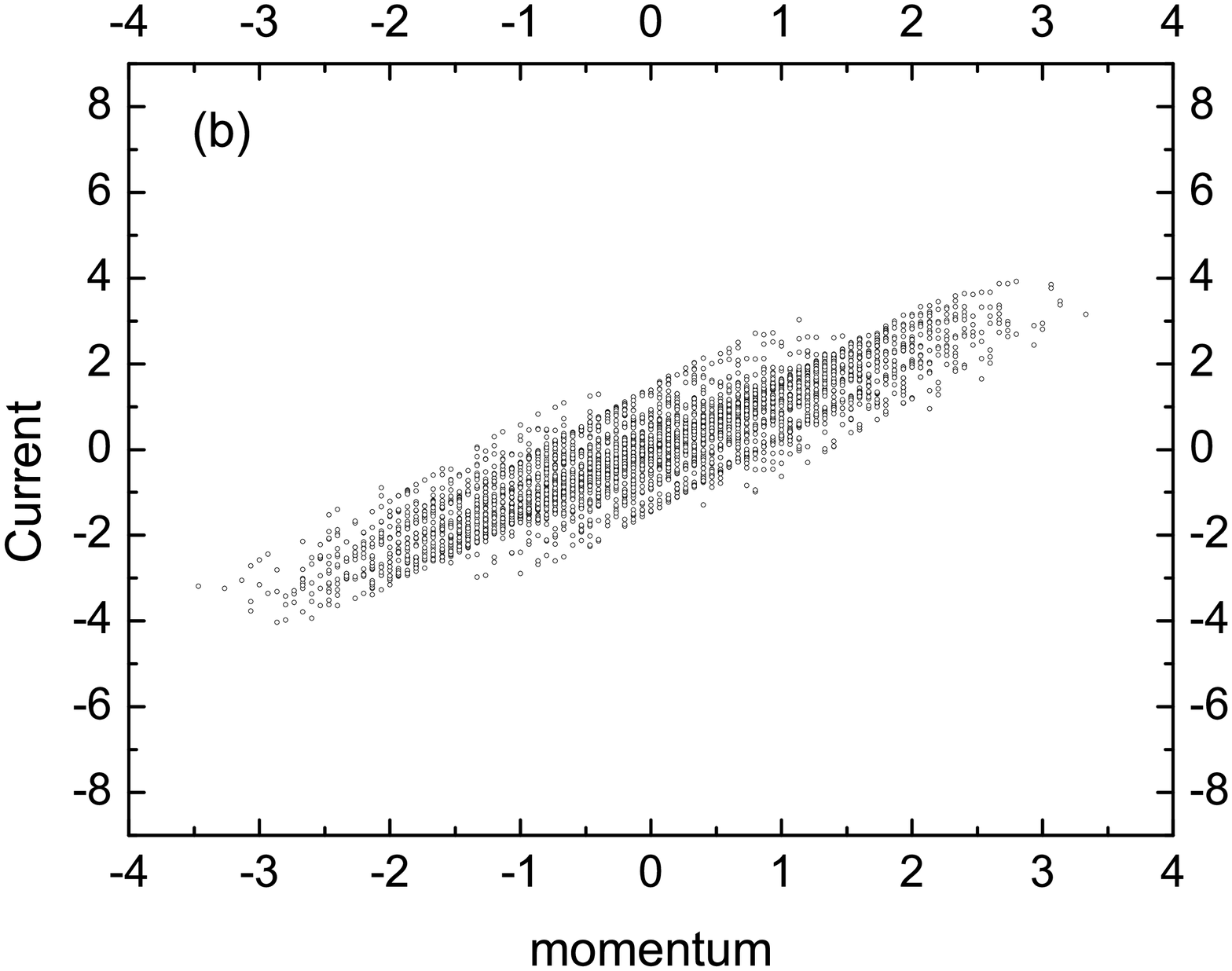}
\includegraphics[scale=0.30]{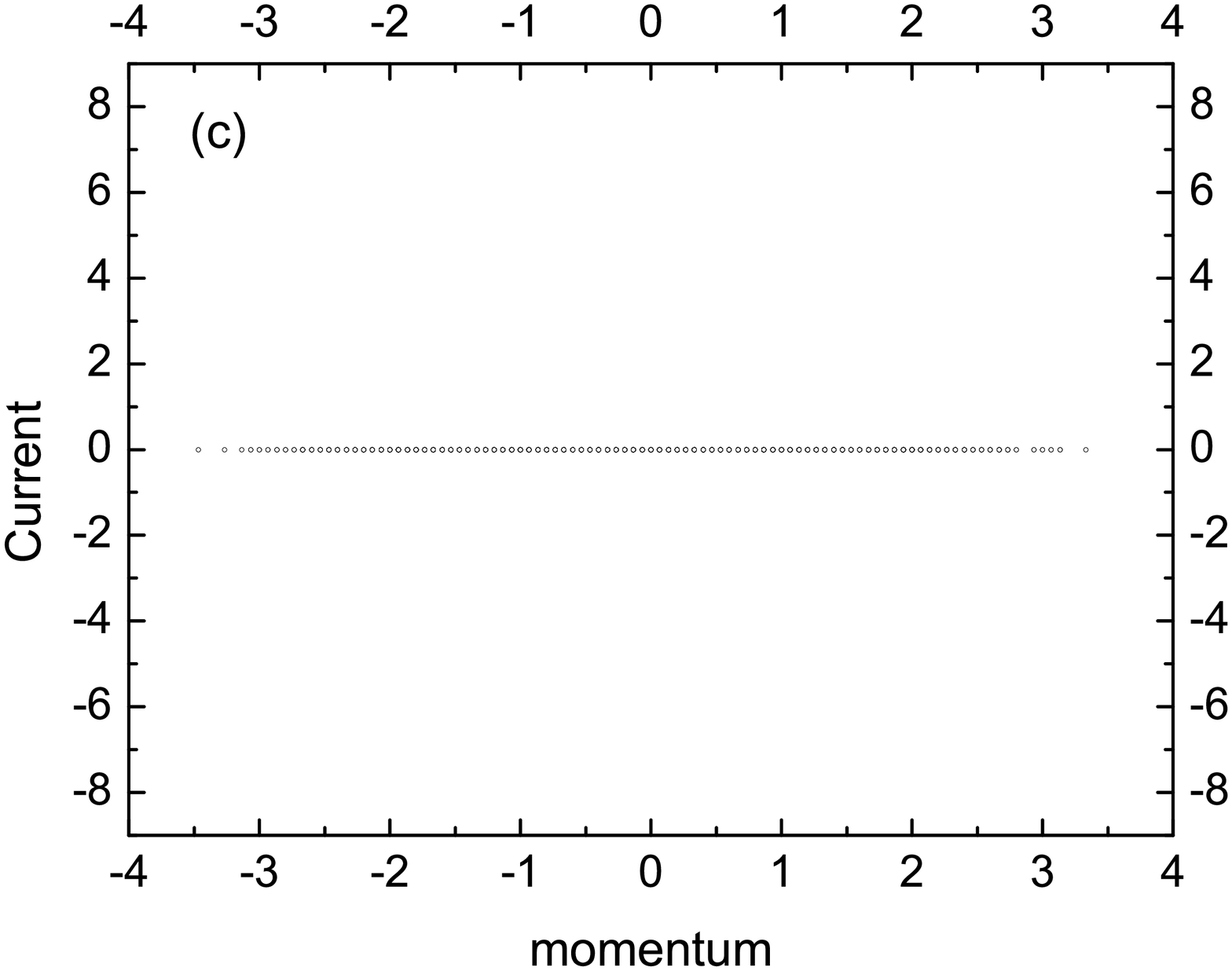}
\includegraphics[scale=0.30]{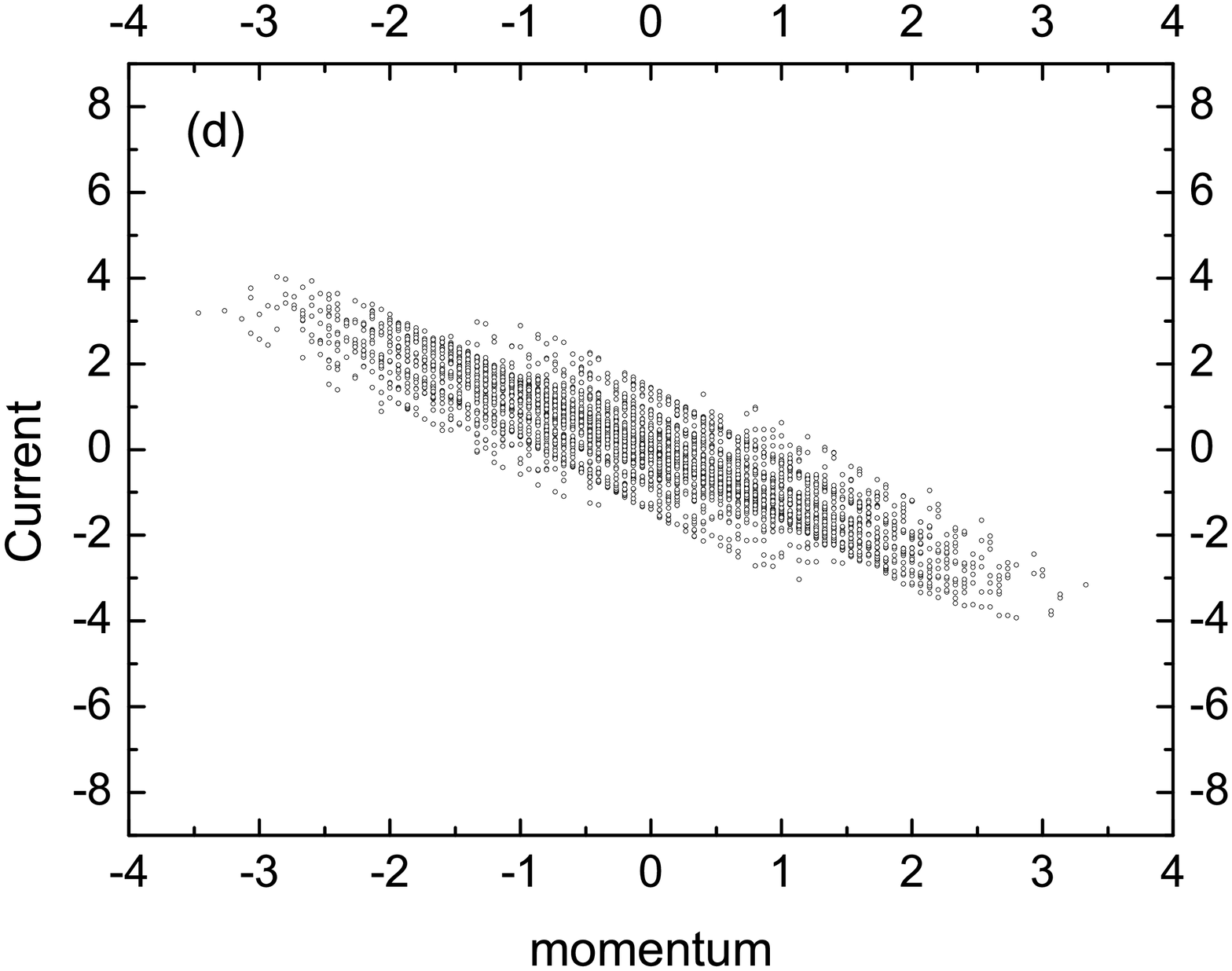}
\includegraphics[scale=0.30]{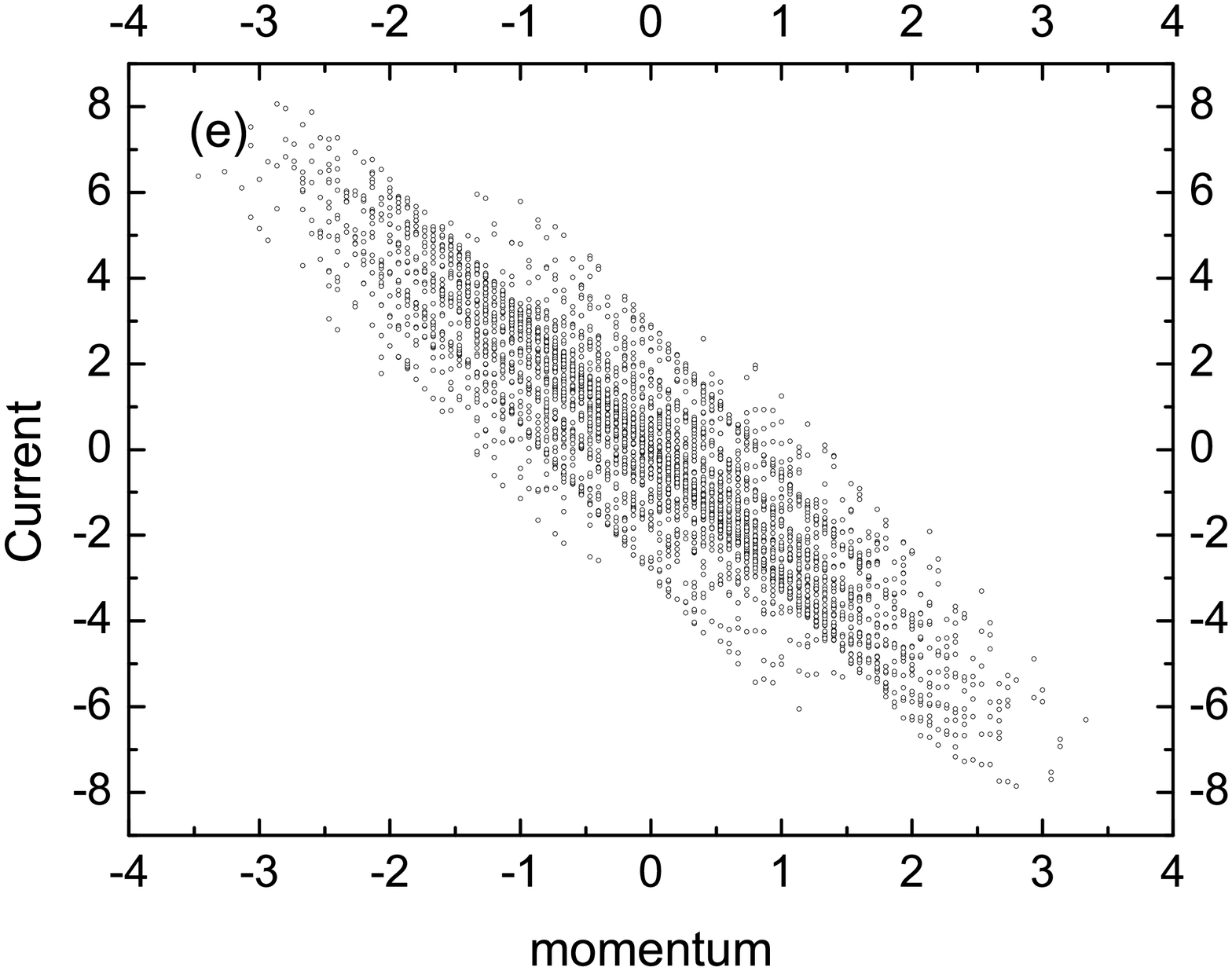}
\caption{\label{fig2}The current spectra of (a) the metallic $S_{\eta}=2;S_{\eta}^{z}=-2;M'=0$  
states, (b) metallic $S_{\eta}=2;S_{\eta}^{z}=-1;M'=0$ states,
(c) half-filling $S_{\eta}=2;S_{\eta}^{z}=0;M'=0$ states, $S_{\eta}=1;S_{\eta}^{z}=0;M'=1$ states,
and $S_{\eta}=S_{\eta}^{z}=0;M'=2$ states, 
(d) metallic $S_{\eta}=2;S_{\eta}^{z}=1;M'=0$ states, and
(e) metallic $S_{\eta}=2;S_{\eta}^{z}=2;M'=0$ states considered in the text for $ U/t=6$ and $N_a=30$.}
\end{figure}

\section{Current spectra of degenerate half-filling and metallic states}
\label{curr-spec}

In order to illustrate that $n=1$ and $n\neq 1$ energy eigenstates whose
energy spectra are degenerate carry when $U/t>0$ zero and finite charge current, respectively, we have 
derived numerically the current expectation value and energy spectra of a set of related energy eigenstates 
with $2S_c^h = 2[S_{\eta}+M'] = 4$ and thus four holes in the $c$ momentum band \cite{1D-03-04}.
The energy spectrum of the simpler energy eigenstates with $2S_c^h = 2[S_{\eta}+M'] = 2$
were studied and plotted in Ref. \cite{GPC-07}. Such states
have two holes in the $c$ momentum band \cite{1D-03-04} and include
three types of $S_{\eta}^z=0,\pm 1$; $S_{\eta} =1$; $M'=0$ states and the
$S_{\eta} = -S_{\eta}^z=0$; $M'=1$ states. Only the charge current spectrum
of the metallic $S_{\eta}^z=-1$; $S_{\eta} =1$; $M'=0$ states was plotted in
Ref. \cite{GPC-07}. On the other hand, the charge currents carried by the
also metallic $S_{\eta}^z=+1$; $S_{\eta} =1$; $M'=0$ states is minus that
of the $S_{\eta}^z=-1$; $S_{\eta} =1$; $M'=0$ states. Moreover, the 
half-filling $\eta$-spin-triplet $S_{\eta}^z=0$; $S_{\eta} =1$; $M'=0$ states
and half-filling $\eta$-spin-singlet $S_{\eta} = -S_{\eta}^z=0$; $M'=1$ states
carry no current in the thermodynamic limit.

Here we consider five types of $S_{\eta}^z=0,\pm 1,\pm 2$; $S_{\eta} =2$;
$M'=0$ states, three types of $S_{\eta}^z=0,\pm 1$; $S_{\eta} =1$; $M'=1$ states, and two types of 
$S_{\eta} = -S_{\eta}^z=0$; $M'=2$ states with two and one occupied BA quantum number $J^{'1}_{\alpha}$ 
and $J^{'2}_{\alpha}$ of Ref. \cite{Takahashi}, respectively. The degenerate energy
spectrum of such states is plotted in Fig. \ref{fig1}. The current spectra of
the energy eigenstates with $2S_c^h = 2[S_{\eta}+M'] = 4$ and $\eta$-spin $S_{\eta} =0$ and $S_{\eta} =2$
are plotted in Fig. \ref{fig2}. The spectra of Figs. 1,2(a) were calculated from the BA for $U/t=6$ and $N_a=30$.
Consistent with the results of this paper, the current of the $n=1$ states plotted
in Fig. \ref{fig2}(c) vanishes. The metallic states whose 
currents are plotted in Figs. \ref{fig2}(a),(b),(d),(e) have the same energy as
these $n=1$ states yet carry finite charge current.

\section{Conclusions}
\label{conclusions}

Recently the finite-energy behavior of correlation functions of 1D correlated systems
\cite{TTF,Glazman,DSF-n1,tj-13} has been found to differ signiÞcantly 
from the linear Luttinger liquid theory predictions \cite{Voit}. Here we have considered
the related problem of the exotic $T>0$ charge transport properties of the half-filled 1D Hubbard model. 
We have shown that for $U/t>0$ its charge stiffness $D (T)$ vanishes for $T>0$ in the thermodynamic limit.  
The corresponding absence of finite-temperature ballistic charge transport  
is an exact result that clarifies a long-standing open problem. 

Whether for the half-filled 1D Hubbard model $\sigma_0 = \lim_{\omega\rightarrow 0}\sigma_{reg} (\omega,T)$ vanishes
or is finite for $T>0$ and $U/t>0$ in the thermodynamic limit is an interesting related open issue that deserves further investigations.\\ \\
{\bf Acknowledgments}\\ \\
We thank H. Johannesson, H. Q. Lin, N. M. R. Peres, and P. D. Sacramento for discussions, the hospitality and support 
of the Beijing Computational Science Research Center, and the
support of the Research Grants Council of Hong Kong under the Project CUHK401108. J. M. P. C. thanks the hospitality 
of the University of Gothenburg and the support by the FEDER through the COMPETE Program, Portuguese FCT both 
in the framework of the Strategic Project PEST-C/FIS/UI607/2011 and under SFRH/BSAB/1177/2011, and
the Swedish Foundation for International Cooperation in Research and Higher Education (Grant. No. IG2011-2028).
S.-J. Gu is supported by the Earmarked Grant Research from the Research 
Grants Council of HKSAR, China under the project CUHK 401212.\\ \\
{\bf References}\\

\end{document}